\documentclass{article}

\usepackage{amssymb}
\usepackage{graphicx}
\usepackage{lineno}
\usepackage{hyperref}
\usepackage{amsmath,amssymb,amsthm,mathrsfs,amsfonts,dsfont} 
\usepackage{fancyhdr}
\usepackage{float}
\usepackage{pdflscape}
\usepackage{bigints}
\usepackage{caption}
\usepackage{subfig}  
\usepackage[T1]{fontenc}   
\usepackage{mathptmx}      

\def\R{\mathbb{R}}

\newcommand{\cov}{\mathbb{COV}}

\usepackage{xcolor}
\usepackage{soul}
\sethlcolor{green}

\newcommand{\revA}[1]{#1} 
\newcommand{\revB}[1]{#1} 

\begin{document}

\title{The ICSCREAM methodology: Identification of penalizing configurations in computer experiments using screening and metamodel -- Applications in thermal-hydraulics}
%
%

\author{A. Marrel,\\
CEA, DES, IRESNE, DER, F-13108, Saint-Paul-lez-Durance, France\\ 
\\
B. Iooss and V. Chabridon\\
EDF R\&D, 6 Quai Watier, 78401, Chatou, France}

\maketitle
\date

\begin{abstract}
In the framework of risk assessment in nuclear accident analysis, best-estimate computer codes, associated to a probabilistic modeling of the uncertain input variables, are used to estimate safety margins. A first step in such uncertainty quantification studies is often to identify \revA{the critical configurations (or penalizing, in the sense of a prescribed safety margin)} of several input parameters (called ``scenario inputs''), \revA{under} the uncertainty on the other input parameters. However, the large CPU-time cost of most of the computer codes used in nuclear engineering, as the ones related to thermal-hydraulic accident scenario simulations, involve to develop highly efficient strategies. This work focuses on machine learning algorithms by the way of the \revA{metamodel-based approach} (i.e., a mathematical model which is fitted on a small-size sample of simulations). To achieve it with a very large number of inputs, a specific and original methodology, called ICSCREAM (Identification of penalizing Configurations using SCREening And Metamodel), is proposed. The screening of influential inputs is based on an advanced global sensitivity analysis tool (HSIC importance measures). A Gaussian process metamodel is then sequentially built and used to estimate, within a Bayesian framework, the conditional probabilities of exceeding a high-level threshold, according to the scenario inputs. The efficiency of this methodology is illustrated on two high-dimensional (around a hundred inputs) thermal-hydraulic industrial cases simulating an accident of primary coolant loss in a pressurized water reactor. For both use cases, the study focuses on the peak cladding temperature (PCT) and critical configurations are defined by exceeding the $90\%$-quantile of PCT. In both cases, the ICSCREAM methodology allows to estimate, by using only around one thousand of code simulations, the impact of the scenario inputs and their critical areas of values.
\end{abstract}

\section{Introduction}\label{sec:Introduction}

In the framework of risk assessment in nuclear accident analysis, best-estimate computer codes are increasingly used to understand, model and predict physical phenomena and, ultimately, estimate safety margins. These codes (also known as ``numerical simulators'') usually take a large number of input parameters driving the phenomenon of interest or related to its physical and numerical modeling. The available information about some of these parameters is often limited or uncertain. The uncertainties come mainly from the lack of knowledge about the underlying physics and about the characterization of the input parameters of the model (e.g., due to the lack of experimental data) \cite{baczha20}. Other additional sources of uncertainty can be considered such as the choice of a particular accident scenario. All these input parameters, and consequently the simulator output, are uncertain. In this context, it is essential to take the uncertainties tainting the results of computer simulations into account \cite{maysor89}. 
This major step for safety studies is referred to as ``uncertainty quantification of numerical models'' \cite{smi14} in the statistical community and is called ``Best-Estimate Plus Uncertainty'' (BEPU) in nuclear safety analysis \cite{wil13,serhop14}.

In this work, we focus on the identification of the penalizing configurations (corresponding to critical values of the output) of specific scenario inputs \revA{(i.e., defining the accident scenario considered)}, \revA{under} the uncertainty of the other inputs. \revA{It is important to notice that the goal is not to perform a ``worst-case'' scenario analysis as often encountered in common nuclear safety practices. The idea here is to look for the inputs' domain leading to a high probability level that the physical variable of interest reaches some large values. Such an approach differs from finding the inputs' values which lead to the maximum of the variable of interest (worst-case analysis).}
\revA{In the uncertainty quantification literature, several families of methods can be found for this particular analysis: on the one hand, it is called ``inversion'' or ``identification of an excursion set'' \cite{picgin10,chebec14} when no uncertainty is considered; on the other hand, it is called ``robust inversion'' \cite{ricbac19} if the uncertainties about other inputs are considered. In the present paper, the proposed methodology rather belongs to the second type of methods. As mentioned in \cite{ricbac19}, the real industrial application tackled in the present paper goes beyond standard canonical problems (e.g., purely optimization or inversion problems) and thus implies to develop a dedicated methodology.} Our study is motivated and guided by the ``Intermediate Break Loss Of Coolant Accident'' (IB-LOCA) safety analysis, based on the numerical simulation of an accident of primary coolant loss in a pressurized water reactor \cite{sansan18}.
In the present paper, a realistic reactor-scale modeling of an IB-LOCA is considered \cite{chalec16,lar19}. This model, relying on the CATHARE2 code \cite{gefant11}, enables to compute thermal-hydraulic transients with respect to a very high number (compared to previous simplified studies \cite{ioomar19}) of input uncertain parameters.

\revB{Standard approaches, related to our objectives and used in the nuclear industry (in particular on the IB-LOCA case), are for example:
\begin{itemize}
    \item the inverse uncertainty quantification methods \cite{men18,baczha19,damgai20} to obtain the probabilistic distributions of some physical model inputs when experimental data are available, 
    \item the Wilks method \cite{wal07,petdau08} to infer high quantiles of the model output variables of interest, 
    \item the PLI method \cite{lar19,ioover21} to detect the most influential inputs to be penalized amongst a large number of variables before applying the Wilks method.
    \item and finally the RIPS (Reduction of the Interval of variation of the Parameters of the Scenario) method, recently published in \cite{lecsau19}, aims to characterize the limiting scenario in a BEPU approach.
\end{itemize}}
\noindent
In particular, RIPS allows to analyze the high-order (or low-order) quantiles of the output cumulative distribution function and determine, for each scenario input, the critical zone within its variation interval. A first issue of this method is that it relies on a quite subjective visual analysis, but its most important drawback is due to the intrinsic complexity in the tuning of the method. 

Our goal is to provide a more robust and automatic methodology which allows to reduce the computational cost (in terms of number of code runs) compared to the RIPS approach. In uncertainty quantification studies, to solve the cost issue, a widely accepted approach consists in approximating the CPU-time expensive computer models by CPU-time inexpensive mathematical functions called ``metamodels'' (or ``surrogate'' models). These metamodels can be based, for instance, on polynomials, neural networks or Gaussian processes \cite{fanli06}.
\revA{The metamodel is built from a set of computer code simulations, and must be as representative as possible of the ``true'' code in the variable domain of the uncertain parameters while having good prediction capabilities.}
Nowadays, metamodels are extensively used in several engineering fields to solve industrial issues as it provides a \revA{multi-purpose} tool \cite{forsob08}: once fitted, the metamodel can be used, in conjunction with the costly computer code, to perform sensitivity analysis, as well as uncertainty propagation, optimization, or calibration studies \revB{see, e.g., \cite{khuwaileh:2017:surrogate,wu:2018:kriging_invuq,whyte:2020:surrogate,christian:2021:dynamic_pra,puppo:2021:failure}).}

An important issue is that the building process of the metamodel remains complex in the case of high-dimensional (e.g., typically several tens of inputs) numerical experiments. In order to build a metamodel in a efficient manner in such cases, \cite{ioomar19} proposed a methodology which combines several \revA{pre-existing} advanced statistical tools: first, an initial space-filling design of experiments; second, a screening step (which aims at detecting influential inputs and non-influential ones) in order to reduce the dimension, and third, \revB{a specific building strategy (based on a sequential inclusion of variables)} of a joint Gaussian process metamodel. Then, the resulting joint Gaussian process metamodel is used to accurately estimate high-order output quantiles. The efficiency of the methodology has been illustrated on a simplified IB-LOCA use case with $27$ inputs and a total budget of $500$ code simulations.

The objectives of the present study are different from those identified in \cite{ioomar19}. In the present paper, the goal is to perform an inversion and thus requires the new proposed statistical methodology, called ICSCREAM (pronounced ``ice-cream''), for ``Identification of penalizing Configurations using SCREening And Metamodel''. Indeed, finding the penalizing configurations in input with respect to critical output values (e.g., typically, a high-order quantile) corresponds to identify specific areas of some specific inputs, while \revA{ignoring} the uncertainty tainting the other ones. Moreover, since the considered use case is a reactor-scale model of an IB-LOCA, the number of input parameters is considerably larger than for standard mock-up cases. Parallel to that, the available simulation budget is fixed (around a thousands simulations). Therefore, a methodological challenge arises in the metamodel building process. 
 
The paper is organized as follows. In Section~\ref{sec:ics_methodo}, the general workflow of the ICSCREAM methodology is first detailed. Two different IB-LOCA use cases and their corresponding datasets are then presented in Section~\ref{sec:use_case}. Thereafter, Sections~\ref{sec:step2} to~\ref{sec:step4} are dedicated to each step of the method: their theoretical foundations and their practical implementation are first detailed before presenting the results obtained on the two datasets. Section~\ref{sec:ccl} provides some conclusions and prospects of this work.
The Appendix (Section \ref{sec:analytic}) provides an additional illustration of the methodology through an analytical test case.



\section{ICSCREAM: a four-step-based methodology}\label{sec:ics_methodo}

For the sake of clarity, a few notations are introduced at this stage. Throughout the rest of this paper, the numerical model (i.e., computer code or simulator) is represented by the following input-output relationship:
\begin{equation}
\mathcal{M}:
\left|
  \begin{array}{rcl}
    \mathcal{X} & \longrightarrow &\mathcal{Y} \\
    \mathbf{X}  & \longmapsto     & Y = \mathcal{M}(\mathbf{X})\\
  \end{array}
\right.
\end{equation}
where the uncertain output variable $Y$ and the $d$ input parameters $\mathbf{X}=(X_1,\dots,X_d)^\top$ belong to some measurable spaces respectively denoted by $\mathcal{Y}$ and $\mathcal{X}\subset\R^d$. As part of the probabilistic approach, the inputs are considered as random variables with probability distributions denoted by $\mathbb{P}_{\mathbf{X}}$ on $\mathcal{X}$ \cite{helton1997uncertainty,oberkampf2001mathematical}. Moreover, among the $d$ uncertain inputs, the $d_{\textrm{pen}}$ scenario inputs (that need to be penalized) are denoted by $\mathbf{X_{pen}} \subset \mathbf{X}$. Notice that no general assumption about independence between the inputs is required here. Only the $\mathbf{X_{pen}}$ are assumed to be independent from $\lbrace \mathbf{X} \setminus \mathbf{X_{pen}} \rbrace$, i.e., from the remaining inputs. Dependency among the $\mathbf{X_{pen}}$ has to be considered conjointly when computing the conditional probabilities for the identification of penalizing configurations (see Section~\ref{sec:step4}). Then, it is supposed that only input-output observations (or realizations) of $\mathcal{M}$ are available. It is therefore assumed that we have a $n$-size sample of inputs and associated outputs denoted by $\left(X_s,Y_s\right)$ where $X_s = \lbrace \mathbf{x}^{(1)},\ldots, \mathbf{x}^{(n)} \rbrace$ with $\mathbf{x}^{(i)} = \left(x_1^{(i)},\ldots,x_d^{(i)}\right)$ denotes the matrix of $n$-size sample locations (also called the ``experimental design''), and $Y_s = \lbrace y^{(1)},\ldots,y^{(n)} \rbrace$ the corresponding outputs observations with $y^{(i)} = \mathcal{M}(\mathbf{x}^{(i)})$.

\begin{figure}[!ht]
\centering
\includegraphics[,height=0.9\textheight]{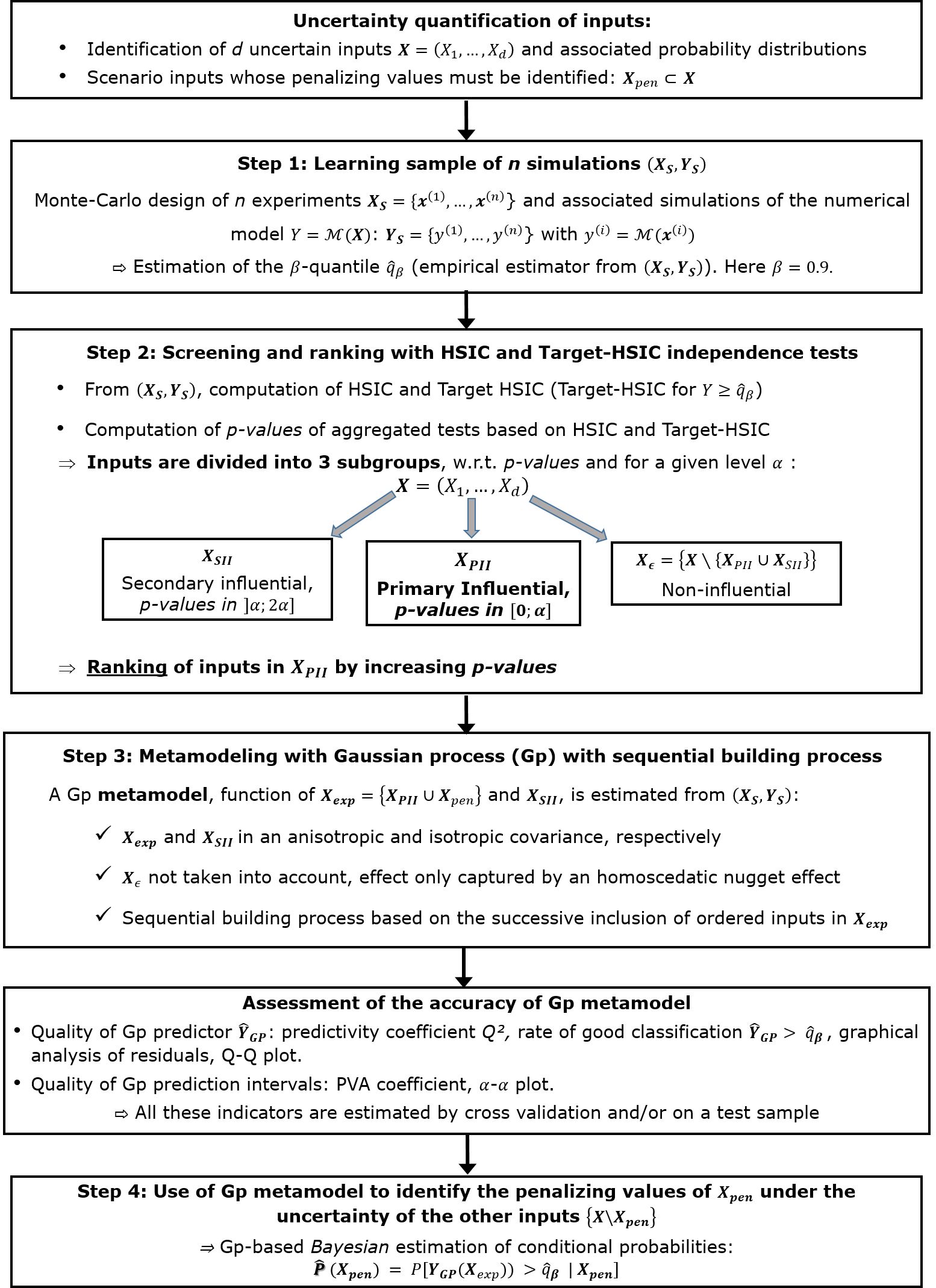}
\caption{\label{fig:Workflow}General workflow of the ICSCREAM methodology.}
\end{figure}

To tackle the problem of large input dimension and complex computer codes, the ICSCREAM methodology combines several statistical techniques \revB{which have been chosen for their relevance regarding the multiple constraints imposed by the problem of interest (i.e., both the dimensionality and the limited size of the learning sample)}. Figure~\ref{fig:Workflow} provides a general workflow of the main steps:
\begin{itemize}
\item \textbf{Step 1: Design of experiments.} Knowing the variation domain of the input variables $\mathbf{X}$, a design of $n$ numerical experiments is firstly performed to obtain the \emph{learning sample} $\left(X_s,Y_s\right)$. To apply statistical theory and methods, this sample must be chosen randomly, from Monte Carlo sampling techniques. The simplest choice is a crude Monte Carlo sample. However, ``stratified'' sampling techniques (such as Latin Hypercube sampling \cite{loh96}) can also be used to ensure a more regular sampling of marginal distributions and a better convergence of statistical estimators. 
\revA{Another particularly relevant design strategy would be to use metamodel-based adaptive designs \cite{forsob08}. This solution would allow to select, in a sequential way, new simulation points (in the inputs' domain) in order to improve the global accuracy of the metamodel or directly the estimation of penalizing configurations.
However, this solution cannot be used in our case: we have to rely on a unique finite batch of random CATHARE simulations (performed at the beginning of the process). The objective is to provide an operational tool for the engineer without any extra interfacing with the simulation code and which is able to deal with a constrained simulation budget.
}
\item \textbf{Step 2: Preliminary screening and ranking with global and target HSIC measures.} From the learning sample, a screening analysis is performed via statistical independence tests based on Hilbert-Schmidt Independence Criterion (HSIC) measures \cite{delmar16b}.
For this, HSIC are considered in global and target versions, ``target'' referring here to the area where the output exceeds a given critical value (i.e., a threshold). From the results of HSIC and target HSIC-based tests, the primary influential inputs (PII) and denoted $\mathbf{X_{PII}}$ are identified. These PII are the most influential inputs regarding the output variability   and can be ranked by decreasing influence. Another set of inputs of lesser (i.e., secondary) influence is gathered in $\mathbf{X_{SII}}$. Finally, all the remaining inputs are considered as global stochastic (i.e., unknown) inputs, denoted by $\mathbf{X_{\epsilon}}$. 

\revB{Unlike in a previous work \cite{ioomar19}, HSIC are also used here in a goal-oriented version \cite{marcha20} to tackle the objective of conditional probability estimation. Moreover, statistical independence tests based on HSIC are used to ensure a more rigorous interpretation in terms of ranking and screening, while taking into account the estimation error related to the limited sample size. Both global and target HSIC must then be combined into aggregated tests to provide a unique screening and ranking.}

\item \textbf{Step 3: Building and validation of Gaussian process (Gp) metamodel.} From the learning sample, a Gp metamodel is built to fit the simulator output $Y$, taking advantage of the results provided by Step 2. More precisely, one considers the ordered explanatory inputs $\mathbf{X_{exp}}$ of the Gp metamodel. Basically, they consist in the $\mathbf{X_{PII}}$ to which we add the inputs that have to be penalized (if not already selected in $\mathbf{X_{PII}}$): $\mathbf{X_{exp}} = \lbrace  \mathbf{X_{PII}} \cup \mathbf{X_{pen}}  \rbrace$. Note that the $\mathbf{X_{SII}}$ are also included as explanatory variables but in a coarser way. Finally, the residual effect of the neglected inputs, merged into $\mathbf{X_{\epsilon}}$, is captured using an additional noise effect.
Concerning the estimation of all the Gp hyperparameters, similarly to \cite{ioomar19}, a sequential process \revA{(which relies on the ranking deduced from the preliminary screening step) is used.} \revB{More precisely, the information provided by HSIC-based tests is adapted in order to address the issue of the large input dimension. The inputs of lesser influence $\mathbf{X_{SII}}$, previously neglected in \cite{ioomar19}, are now injected at the end of the sequential process to reduce the risk of mis-selection (which is common with such sparse data), while limiting the training efforts of the Gp metamodel.} Finally, the accuracy and prediction capabilities of the Gp metamodel are controlled, either on a test sample (if available) or by cross-validation based on the learning sample.
\item \textbf{Step 4: Use of the metamodel for the identification of penalizing configurations}. 
Once built, the Gp can be evaluated in prediction in a very intensive way \revB{(e.g., from $10^5$ to $10^6$ evaluations)}.
Usually, a Gp metamodel can be used for a quantitative sensitivity analysis \cite{davgam21} based on the output variance decomposition (e.g., estimation of Sobol' indices), as well as uncertainty propagation (e.g., estimation of high-/low-order quantiles or probabilities), leading in all cases to a large gain of computation time (see \cite{ioomar19} for an example in nuclear engineering). Here, the Gp metamodel is used to estimate, within a Bayesian framework, the conditional probabilities of exceeding the critical quantile value, \revA{as a function of all the inputs that need to be penalized and ignoring the other uncertain ones. Thus, if the penalized inputs are independent, one will only have to compute a conditional probability per input and, in the case of dependent penalized inputs, this probability will turn into a probability hypersurface.}

\revB{Again, the final purpose of the present paper differs from the one in \cite{ioomar19}: some modifications in the methodology were required, such as the use of target HSIC measures. Moreover, as it will be described in the next section, the datasets treated in the present work go far beyond by considering much more complex reactor-scale datasets with around hundred inputs (in contrast to the simplified mock-up use case with less than thirty inputs).}
\end{itemize}

\revA{As stated before, the motivations for such a methodology are, before anything else, driven by an industrial application illustrated by two use cases described in the following section. However, it is important to notice that this methodology follows a logical modular sequence which makes it rather generic (in the sense that it can be applied to a large class of problems). To emphasize this, an illustration through an analytical test case is provided in Appendix (Section \ref{sec:analytic}).}

\section{Description of the thermal-hydraulic use cases and datasets}
\label{sec:use_case}
In support of regulatory work and nuclear power plant design and operation, safety analysis considers the so-called ``loss of coolant accident'' which takes into account a double-ended guillotine break with a specific size piping rupture. More precisely, one considers here an intermediate break loss of coolant accident (IB-LOCA) \cite{sansan18} \revB{in a three-loops pressurized water reactor from the French fleet\footnote{The input data considered for this case, and thus the obtained PCT, do not correspond to actual industrial values.}}.
The output variable of interest is the maximal peak cladding temperature (PCT) of the hot rod during the accident transient. The reactor coolant system minimum mass inventory and the PCT are obtained shortly after the beginning of the accumulators’ injection \cite{chalec16}. The IB-LOCA transients are simulated using the thermal-hydraulic system code (two phase flow six equations) CATHARE2 \cite{gefant11}, jointly developed by CEA, EDF, Framatome and IRSN. Around a hundred inputs (vector $\mathbf{X}$) are considered as uncertain and can be split into three different types \cite{lar19}:
\begin{itemize}
    \item Type $1$: the boundary and initial conditions whose probability distributions are \revB{supposed to be known} (as uniform or normal distributions);
    \item Type $2$: the model parameters (e.g., models related to two-phase flow hydraulics, models associated to heat transfer and models describing the clad behavior). Their distributions can be obtained from experimental data, expert knowledge or recovered by \revA{assimilation} of experimental databases \cite{baczha20}. This leads to uniform, log-uniform, normal or log-normal distributions;
    \item Type $3$: the scenario parameters \revB{(see their description below)} which cover some variability between minimal $m$ and maximal $M$ bounds \revB{(they follow a uniform distribution that will be noted $U[m,M]$ in the following)}.
\end{itemize}

All the inputs of types $1$ and $2$ are independent. 
\revB{Their detailed list, as well as their probability distribution and range of variations, are not given due to industrial confidentiality.}
The scenario inputs (type $3$) are independent with those of the two other groups but some dependency might exist among them. Usually, the inputs corresponding to type 3 have to be taken at their worst-case values (corresponding to the maximal value which can be reached by the PCT) for \revB{safety demonstration} \cite{chalec16,lecsau19,lar19}. In practice, these worst-case values are unknown and only a domain of variation of each input is given. \revA{Thus, the idea is to find the penalizing values for these scenario inputs.}

\begin{figure}[!ht]
\centering
\includegraphics[width=0.48\textwidth]{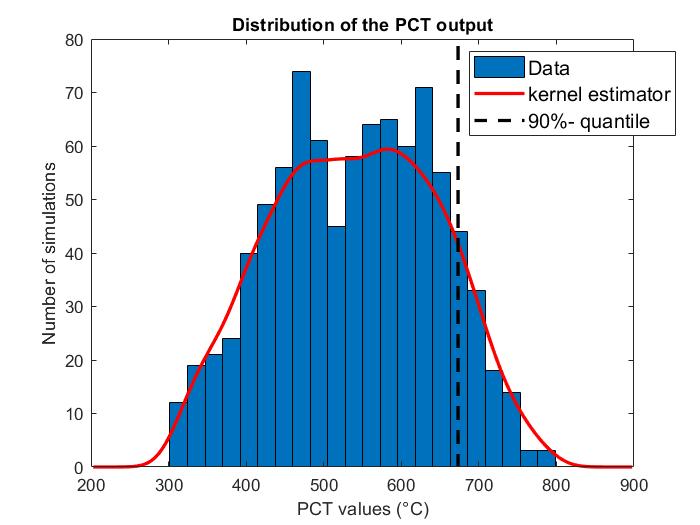}
\includegraphics[width=0.48\textwidth]{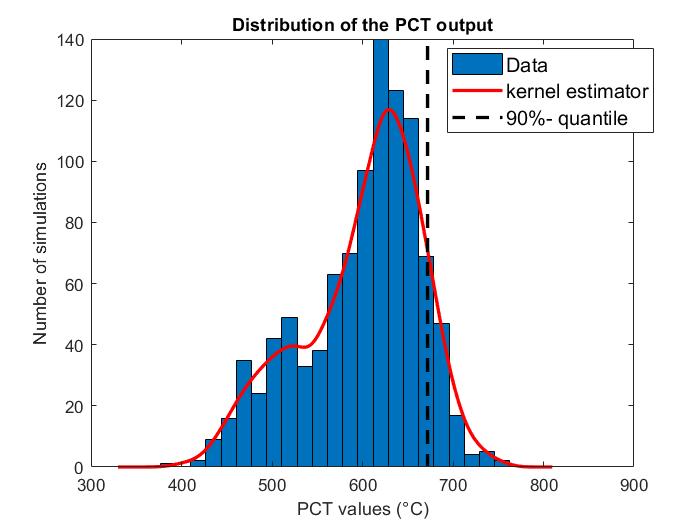}
\caption{Histogram of the PCT from the learning sample for IB-LOCA$_1$ (left) and IB-LOCA$_2$ (right) datasets. The estimated $90\%$-quantiles are indicated by a black dotted line.}\label{fig:hist}
\end{figure}

The following use cases are considered: 
\begin{itemize}
    \item Use-case $\#1$ which corresponds to dataset IB-LOCA$_1$ with $d=96$ uncertain inputs, including $d_{\textrm{pen}}=2$ inputs which have to be penalized: the size of the break, denoted by $X_{92}$ in the dataset \revB{($X_{92} \sim U[3,4.2]$ inches)}, and the stopping time of the primary pumps, $X_{94}$. \revB{$X_{94}$ follows a uniform distribution with a range of variation depending on the size of the break (globally included in $[500,1200]$ s)}.
    It is a small IB-LOCA case without ``loss of offsite power'' (LOOP) including a late main coolant pump coast down;
   \item Use-case $\#2$ which corresponds to dataset IB-LOCA$_2$ with $d=97$ uncertain inputs, including $d_{\textrm{pen}}=10$ independent scenario inputs: hot rod peaking factor elevation \revB{($X_1 \sim U[2.4,3.2]$ m)}, hot rod burn-up  \revB{($X_{14} \sim U[515,59000]$ MWj/t)}, \revB{discharge line accumulator of the intact loop ($X_{34} \sim U[800,1900]$ m$^{-4}$), discharge line accumulator of the broken loop ($X_{35} \sim U[800,1900]$ m$^{-4}$), liquid accumulator enthalpy ($X_{36} \sim U[33544,213105]$ J/kg), temperature of safety injection ($X_{90} \sim U[7,50]$ $^\circ$C), weight factor on flow rate of safety injection $1$ ($X_{91} \sim U[-1,1]$), tube plugging of steam generator of the intact loop ($X_{92} \sim U[0,0.09]$),  weight factor on flow rate of the emergency feedwater system of steam generator ($X_{93} \sim U[-1,1]$), temperature of the emergency feedwater system of steam generator ($X_{94} \sim U[7,55]$  $^\circ$C).}
It is a $15''$ IB-LOCA case with LOOP. 
\end{itemize}
Note that the numbering of inputs is not the same for both use cases. Monte Carlo samples of $n_1=889$ and $n_2=1000$ CATHARE2 simulations are available for IB-LOCA$_1$ and IB-LOCA$_2$ use cases respectively. For both, the inputs are drawn according to their prior probability distributions. The histograms of the obtained values for the output of interest, namely the PCT, are given in Figure~\ref{fig:hist} (temperature is in $^{\circ}$C). A kernel density estimation plot \cite{par62} is also added to provide a graphical illustration of the probability density function. Note that the number $n$ of simulations is a compromise between the CPU time required for each simulation and the number of inputs. For uncertainty propagation and metamodel-building purpose, some rules of thumb propose to choose $n$ at least as large as ten times the input dimension $d$ \cite{loesac09,marioo08}.

\begin{figure}[!ht]
\centering
\includegraphics[width=0.48\textwidth]{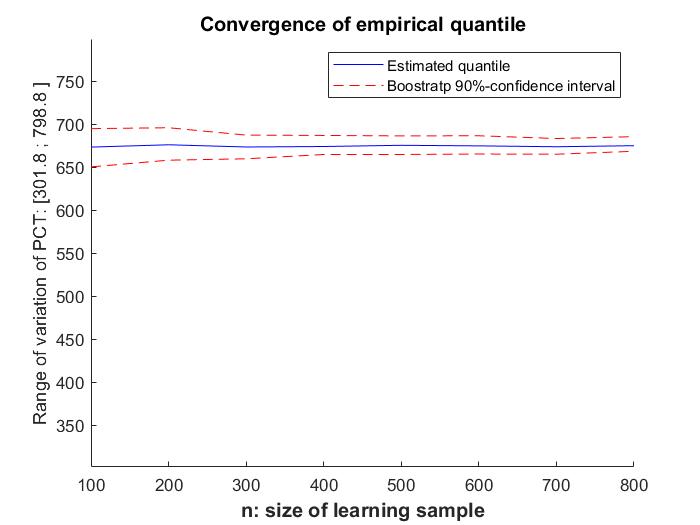}
\includegraphics[width=0.48\textwidth]{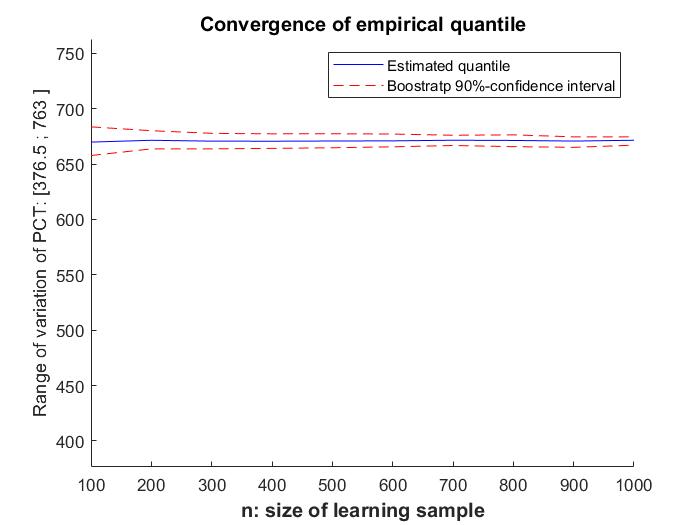}
\caption{\revB{Convergence plots of the empirical estimator of $90\%$-quantile, according to the learning sample size $n$, for IB-LOCA$_1$ (left) and IB-LOCA$_2$ (right).}}\label{fig:MDTE_q90_CV}
\end{figure}

\revB{For the sake of clarity, one reminds that the $\beta$-order quantile (or $\beta$-quantile) is defined by $q_{\beta}~=~\inf \{ y \in \mathbb{R} ~|~ \beta \leq F_{Y}(y) \}$.} From the learning sample, the empirical $90\%$-quantile of PCT is estimated to $\widehat{q}_{0.9}^{(1)} = 673.18^\circ$C (resp. $\widehat{q}_{0.9}^{(2)} =671.85^\circ$C) for IB-LOCA$_1$ (resp. IB-LOCA$_2$). As aforementioned, the ICSCREAM methodology aims to identify the values of the inputs $\mathbf{X_{pen}}$ which yield to a high probability of exceeding this quantile. \revB{Figure~\ref{fig:MDTE_q90_CV} shows the convergence plots of the estimated empirical quantile, from bootstrap method ($y$-axes are scaled to the range of variation of PCT). In addition, the coefficients of variation of $\widehat{q}_{0.9}$ are estimated to be less than $1\%$ ($0.76\%$ and $0.38\%$ for IB-LOCA$_1$ and IB-LOCA$_2$, respectively). The convergence of other statistics of PCT distribution is also verified with a coefficient of variation lower than $1\%$ for the mean and around $2\%$ for the standard deviation.}

\begin{figure}[!ht]
\centering
     \includegraphics[width=0.32\textwidth]{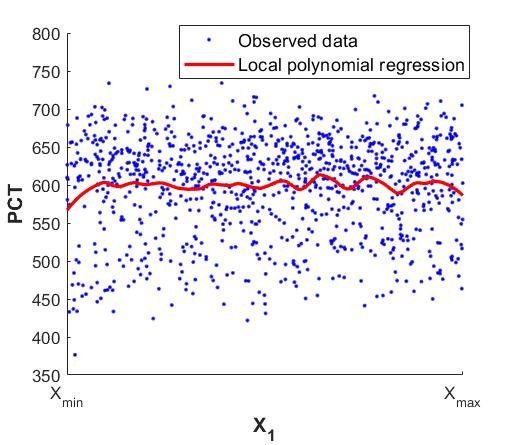}
     \includegraphics[width=0.32\textwidth]{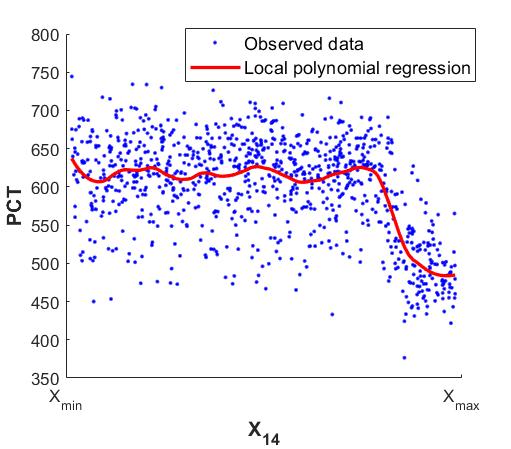}
     \includegraphics[width=0.32\textwidth]{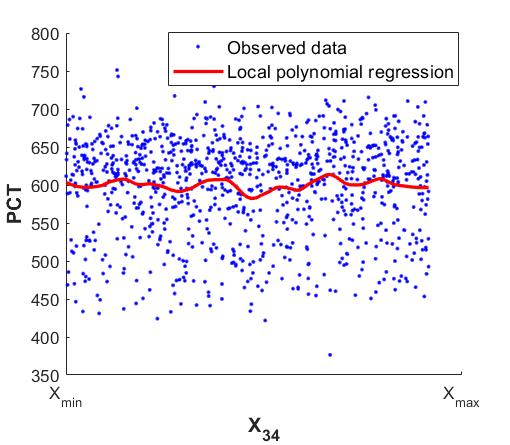}\\
     \includegraphics[width=0.32\textwidth]{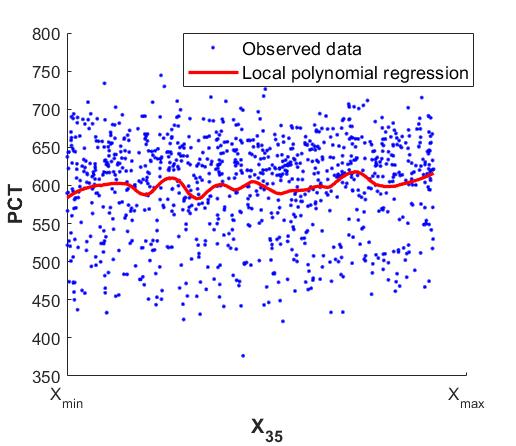}
     \includegraphics[width=0.32\textwidth]{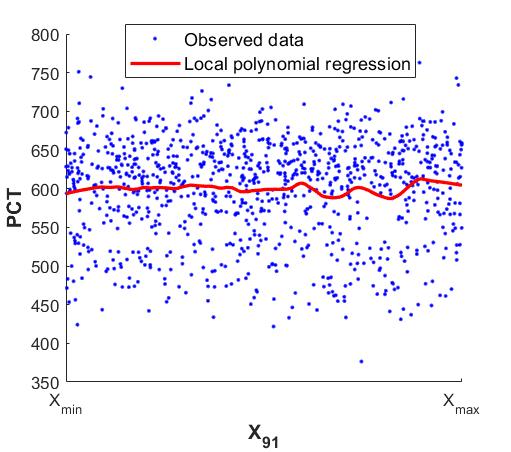}
     \includegraphics[width=0.32\textwidth]{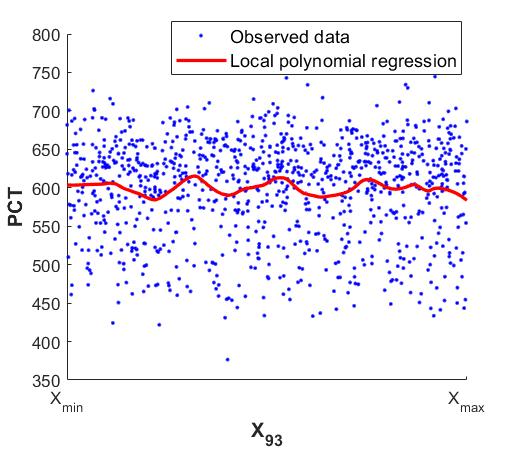}
\caption{\label{fig:MDTE_scatter_scenario}IB-LOCA$_2$: scatter plots of the learning sample with local polynomial regression of the PCT according to several scenario inputs.}
\end{figure}

To illustrate the complexity of fitting the output according to such a large number of input, some scatter plots of the PCT (for IB-LOCA$_2$ dataset) with respect to some scenario inputs are displayed in Figure~\ref{fig:MDTE_scatter_scenario}. In this case, only one scenario input, namely $X_{14}$ (hot rod burn-up), seems to have a clearly detectable influence on the PCT.

\section{Step 2: Screening and ranking with HSIC-based independence tests}\label{sec:step2}

\subsection{Theoretical and methodological details}

An initial screening is performed directly from the learning sample in order to identify the primary and secondary influential inputs (resp. $\mathbf{X_{PII}}$ and $\mathbf{X_{SII}}$ in Figure~\ref{fig:Workflow}). The objective is twofold: first, screening the inputs to reduce the dimension before the metamodeling step and second, ranking them by decreasing order of influence. To achieve it, we use the HSIC importance measures introduced by \cite{grebou05} and theoretically built upon cross-covariance operators in reproducing kernel Hilbert spaces (RKHS). \revB{In a nutshell, the underlying idea of HSIC measures is to capture the influence of an input $X_i$ on the output $Y$ through the measure of their statistical dependence: the more ``correlated'' (in a specific sense that will be explained further) the two variables are, the more influential $X_i$ is.} As highlighted by \cite{dav15} or \cite{delmar16b}, HSIC measures show many advantages for global sensitivity analysis purposes \cite{davgam21}. Some of them are briefly recalled hereafter.
First, HSIC amounts to considering covariance between ``feature'' functions (or transformations) applied to each input-output couple of variables (i.e., an input $X_i$ and the output $Y$). The set of candidate functions (possibly nonlinear) which can be applied is defined by a functional space and an associated kernel. Such a space can be of infinite dimension and thus allows to capture a very broad spectrum of forms of dependency between $X_i$ and $Y$. The HSIC, which is defined as the Hilbert-Schmidt (HS) norm of the cross-covariance operator (denoted by $\cov(\cdot)$), somehow ``summarizes'' the set of covariances between feature functions:
\begin{equation*}
\mathrm{HSIC}(X_i,Y) = || \cov_{X_i Y} ||_{\textrm{HS}} = \sum \limits_{l,m} |  \cov(u_l(X_i),v_m(Y)) |^2,
\end{equation*}
where $(u_l)_{l \geq 0}$ and $(v_m)_{m \geq 0}$ are orthonormal bases of the RKHS associated to $X_i$ and $Y$ respectively.

Furthermore, the so-called ``kernel trick'' allows to get rid of an explicit expression of the features (see \cite{gretton2005measuring} for more details). Thus, HSIC can be directly expressed as a linear combination of expected values of kernels and estimated in a very simple way at rather low cost (typically, a few hundred simulations). By denoting $k_i$ and $k$ the kernels of the two RKHS respectively associated to $X_i$ and $Y$, it can be shown (see \cite{gretton2005measuring}) that the HSIC measure can be
easily estimated by:
\begin{equation}\label{eq:HSIC_Estimator_Trace}
\widehat{\mathrm{HSIC}}(X_i,Y) = \frac{1}{n^2} \mathrm{Tr}( L_i H L H), 
\end{equation}
where $\mathrm{Tr}(\cdot)$ is the trace operator, $L_i$ and $L$ are two Gram matrices defined by $L_i = (  k_i( X_i^{(l)},X_i^{(m)} )  )_{1 \leq l,m \leq n}$ and $L = (  k( Y^{(l)},Y^{(m)} )  )_{1 \leq l,m \leq n}$ with
$(X_i^{(m)},Y^{(m)})_{1 \leq m \leq n}$ a $n$-sample of $(X_i,Y)$, and $H = ( \delta_{lm} - 1/n)_{1 \leq l,m \leq n}$ with $\delta_{lm}$ the Kronecker operator. 

Another key aspect relates to the choice of kernels. Characteristic kernels (such as, among others, the Gaussian one), allow to fully characterize the independence between the input $X_i$ and the output $Y$ with HSIC. This leads to the following fundamental property: the nullity of HSIC is equivalent to independence between $X_i$ and $Y$. Therefore, one can build dedicated statistical independence tests for screening purposes \cite{delmar16b}. The Gaussian kernel, widely used for real variables, is defined by $\ell_\lambda(z,z') = \exp\left(- \frac{\lambda }{2} ( z- z' )^2 \right)$ and parametrized by the bandwidth parameter $\lambda$, often set at $\lambda = 1/\sigma_z^2$ with $\sigma_z^2$ the empirical variance of the sample of $Z$.

Thus, for a given input $X_i$, statistical HSIC-based tests aim at testing the null hypothesis ``$\mathcal{H}_0^{(i)}$: $X_i$ and $Y$ are independent'', against its alternative ``$\mathcal{H}_1^{(i)}$: $X_i$ and $Y$ are dependent''. The significance level (i.e., the probability of rejecting the null hypothesis $\mathcal{H}_0$ when it is true) of this test is hereinafter noted $\alpha_{\textrm{test}}$ and usually set at $5\%$ or $10\%$. Depending on the size $n$ of the considered sample, several versions of HSIC-based tests are available: asymptotic versions (i.e., for large $n$) based on an approximation with a Gamma law \cite{gretton2008kernel}, spectral extensions and permutation-based versions \cite{delmar16b} or adaptive strategies \cite{elamar21} for non-asymptotic cases (i.e., small $n$). 
The p-value is the probability of obtaining HSIC values
either equal to, or larger than, the observed HSIC (estimated from the learning sample) assuming that $\mathcal{H}_0$ true (i.e., that $X_i$ and $Y$ are independent). Such a p-value is used to decide whether to reject (or not) the null hypothesis. Beyond the screening task, the p-value of independence tests can be quantitatively interpreted for ranking the PII, since it can be viewed as a ``margin'' from independence. The lower the p-value, the stronger $\mathcal{H}_0^{(i)}$ is rejected and the higher the influence of $X_i$.
 
Furthermore, remember that the final objective of ICSCREAM is to identify the penalizing configurations and, more precisely, to accurately identify the critical input areas where the PCT exceeds $\widehat{q}_{0.9}$ (i.e., $Y>\widehat{q}_{0.9}$). Consequently, we consider an additional ``target'' sensitivity analysis based on so-called ``target HSIC'' (T-HSIC) indices \cite{marcha20}. Applied here, T-HSIC and associated independence tests aim at measuring the influence of an input $X_i$ over the occurrence of the event $\{Y> \widehat{q}_{0.9}\}$. T-HSIC is built with specific kernels and a weight function can be used in order to cope with the possible loss of information around the threshold (see \cite{marcha20} for further details). 

Finally, both HSIC and T-HSIC tests' results are aggregated to extract a unique screening and ranking: heuristic choices \cite{chamar21} or more robust alternatives (e.g., Bonferroni's correction, p-values combination as inspired from \cite{hearub18}) can be used. More details on Step 2 of the ICSCREAM methodology are given in \cite{chamar21}.

\subsection{Application on IB-LOCA uses-cases}

\begin{figure}[h!]
\begin{center}
	\subfloat[\scriptsize Global-SA-oriented screening.]{\includegraphics[width=0.5\textwidth]{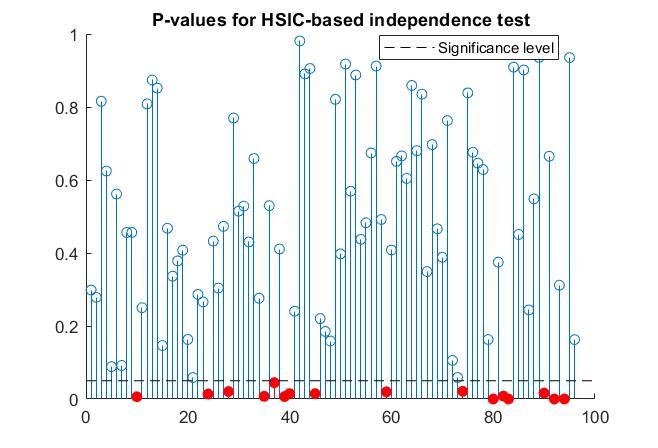}
	\label{fig:plot_pval_GSA_IBLOCA1}}
	~
	\subfloat[\scriptsize Target-SA-oriented screening.]{\includegraphics[width=0.5\textwidth]{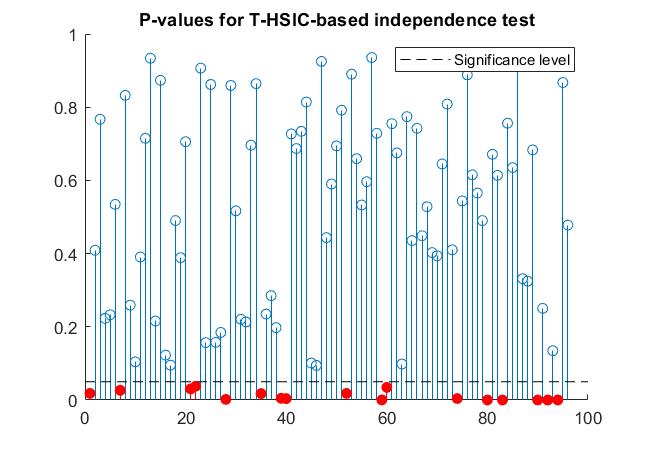}
	\label{fig:plot_pval_TSA_IBLOCA1}}		
\end{center}
\caption{IB-LOCA$_1$ use case: p-values of HSIC and T-HSIC-based independence tests computed from the learning sample. The level $\alpha_{\textrm{test}} = 5\%$ is represented in black dotted line.}\label{fig:plot_pval_IBLOCA1}
\end{figure}

\begin{figure}[h!]
\begin{center}
	\subfloat[\scriptsize Global-SA-oriented screening.]{\includegraphics[width=0.48\textwidth]{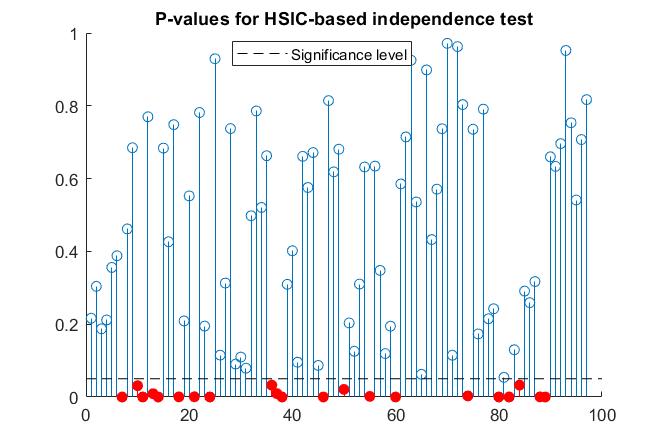}
	\label{fig:plot_pval_GSA_IBLOCA2}}
	~
	\subfloat[\scriptsize Target-SA-oriented screening.]{\includegraphics[width=0.48\textwidth]{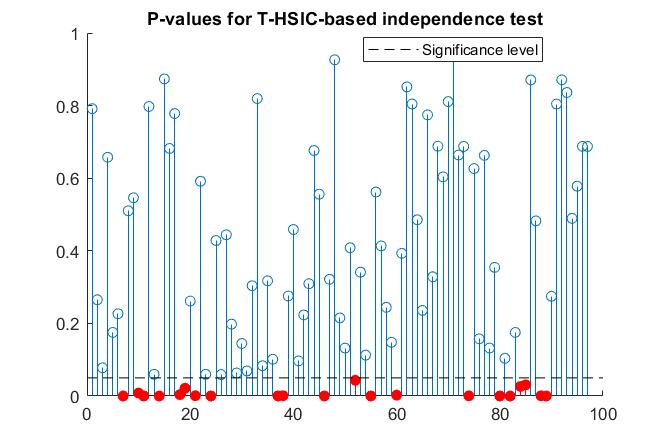}
	\label{fig:plot_pval_TSA_IBLOCA2}}		
\end{center}
\caption{IB-LOCA$_2$ use case: p-values of HSIC and T-HSIC-based independence tests computed from the learning sample. The level $\alpha_{\textrm{test}} = 5\%$ is represented in black dotted line.}\label{fig:plot_pval_IBLOCA2}
\end{figure}

From the learning sample of the two IB-LOCA use cases, HSIC-independence tests are applied with a permutation-based approach for the estimation of p-values. The number of permutations is optimized using the algorithms proposed by \cite{elamar21}. The results are given in Figures~\ref{fig:plot_pval_IBLOCA1} and~\ref{fig:plot_pval_IBLOCA2}. They illustrate the screening process based on p-values: variables whose p-values are below the level $\alpha_{\textrm{test}} = 0.05$ correspond to influential variables (red dots). 
In both use cases, a little bit more than twenty inputs are identified as influential (i.e., p-values < $\alpha_{\textrm{test}}$) during the screening step following the aggregation of global and target HSIC-based tests. About five additional inputs of lesser influence are added in $\mathbf{X_{SII}}$ (p-values in $]\alpha_{\textrm{test}} ; 2\alpha_{\textrm{test}}]$).

\revA{Concerning} the ranking now, for IB-LOCA$_1$, the two inputs to be penalized are the most influential: the stopping time of primary pumps ($X_{94}$) being the most influential one, followed by the break size ($X_{92}$). Then five other inputs are strongly influential, namely the upper plenum and core interfacial friction, core interfacial friction, the hot spot for the hot rod and two inputs relative to accumulators. A group of thirteen other variables of lower influence is also selected by global HSIC-tests. Similar results are obtained with T-HSIC-based tests, except that two additional inputs, namely the diphasic degradation law of pumps and the residual power, are selected as very influential.

For IB-LOCA$_2$, a hard core of around sixteen variables are identified by all the tests with zero p-values. The most influential are the burn-up of the hot rod ($X_{14}$), the heat transfer coefficient of the wall-steam exchange of the hot rod ($X_{46}$) and the bubbly-slug in the core during blowdown phase, related to the interfacial friction ($X_{80}$).
To these sixteen variables, five variables are added (p-values between $0$ and $0.05$).
Regarding the final objective of penalizing the ten scenario parameters, only two of them, namely X$_{14}$ and, with a lesser influence, the liquid accumulator enthalpy (X$_{36}$), are part of the inputs selected by global and target sensitivity analysis. No dependency between the PCT and one of the eight other scenario parameters is detected. Finally, seven additional inputs are added in $\mathbf{X_{SII}}$. 

Note that some convergence studies (with respect to the sample size) not presented here show that the total number of inputs in $\lbrace \mathbf{X_{PII}} \cup \mathbf{X_{SII}}\rbrace$ is relatively stable: inputs in $\mathbf{X_{SII}}$ passing through the group of $\mathbf{X_{PII}}$ as $n$ (and so, the detection capability of the tests) increases. This suggests the gradual and ongoing convergence of the method, as well as its robustness. It also highlights the interest of considering this group of $\mathbf{X_{SII}}$ when screening results are not converged yet. 

\section{Step 3: Gaussian process metamodeling}\label{sec:step3}

\subsection{Parametric choices for the Gaussian process and estimation of the hyperparameters}

The second step of the ICSCREAM methodology consists in building a metamodel, whose aim is to fit the simulator output $Y$ (here, the PCT), based on the learning sample $\left(X_s,Y_s\right)$. To do so, we use a similar approach than the one used in \cite{ioomar19}, based on an homoscedastic (non-interpolating) Gp metamodel. The reader can refer to \cite{raswil06} for a detailed review about Gp metamodel.

As introduced in Section~\ref{sec:ics_methodo}, the inputs variables are divided into three groups at the end of Step 2: 
\begin{itemize}
\item the ordered explanatory inputs $\mathbf{X_{exp}}$: this group is made up of the $\mathbf{X_{PII}}$ to which the inputs to be penalized $\mathbf{X_{pen}}$ are added (if not selected in $\mathbf{X_{PII}}$ yet). In this last case, inputs to be penalized are added at the end, in an arbitrary order, or guided by expert judgment on their supposed importance. We obtain $\mathbf{X_{exp}} = \lbrace  \mathbf{X_{PII}} \cup \mathbf{X_{pen}}  \rbrace$;
 \item a second group of inputs $\mathbf{X_{SII}}$ of much less influence;
 \item the remaining and neglected inputs merged into $\mathbf{X_{\epsilon}}$ with $\mathbf{X_{\epsilon}} = \lbrace  \mathbf{X}\setminus \lbrace \mathbf{X_{exp}} \cup \mathbf{X_{SII}}  \rbrace \rbrace$.
\end{itemize}

The output is thus redefined by $ Y = \mathcal{M}(\lbrace \mathbf{X_{exp}} \cup \mathbf{X_{SII}}  \rbrace, \mathbf{X_\varepsilon)}$ and the metamodeling process is focused on fitting the random variable $Y|\lbrace \mathbf{X_{exp}} \cup \mathbf{X_{SII}}  \rbrace$\footnote{\label{Yrandom}$Y|\lbrace\mathbf{X_{exp}}\cup \mathbf{X_{SII}} \rbrace$ (i.e., $Y$ knowing $\lbrace \mathbf{X_{exp}} \cup \mathbf{X_{SII}}  \rbrace$) is still a random variable as its value depends on the uncontrollable random vector $\mathbf{X}_\varepsilon$.}. In other words, only the inputs in $\lbrace \mathbf{X_{exp}} \cup \mathbf{X_{SII}}  \rbrace$ are considered as the explanatory inputs of the Gp. Basically, the Gp is built to approximate the expected value $\mathbb{E}(Y|\lbrace \mathbf{X_{exp}} \cup \mathbf{X_{SII}}  \rbrace)$. The residual effect of the other inputs (merged into $\mathbf{X}_\varepsilon$) is captured using an additional ``nugget'' effect.
Borrowed from geostatistics, a nugget effect assumes an additive white noise effect and relaxes the interpolation property of the Gp metamodel. It can be assumed to be either constant (homoscedastic) or dependent on the variables  (heteroscedastic). Note that contrary to \cite{ioomar19}, a simple Gp metamodel with an homoscedastic nugget effect is estimated, since fitting a joint Gp with heteroscedastic nugget is neither relevant nor realistic given the high dimensionality of the problem and the rather small sample size. Concerning the parametric choices for the Gp, a constant trend or a one-degree polynomial with a selection algorithm (e.g., the Lasso algorithm \cite{efrhas04}) is considered. Moreover, a stationary anisotropic Mat\'{e}rn $5/2$ tensorized covariance function is used for the inputs in $\mathbf{X_{PII}}$ while $\mathbf{X_{SII}}$ are joint in a stationary isotropic Mat\'{e}rn $5/2$ covariance function (still tensorized with the one of $\mathbf{X_{PII}}$).

\revA{Note that, in order to have a robust and sufficiently automated methodology, some parameters should be fixed. For instance, we recommend in practice to use a Mat\'{e}rn covariance function which is quiet popular in the machine learning community as it covers a large spectrum of applications. Moreover, the resulting differentiability properties of the Gp metamodel (one or two times mean-square differentiable for Matérn $3/2$ and $5/2$, respectively) is a reasonable compromise given the high-dimensional input space, the sparsity of data and the potentially low regularity of the code output. Anyway, various covariance functions can be tested in order to select the best one from the analysis of validation metrics (as done here).} 

\revA{All the Gp hyperparameters (i.e., covariance parameters) are estimated by maximum likelihood estimation on the learning sample. More precisely, a sequential process which relies on the ranking deduced from the screening step is used} \revB{(see \cite{welbuc92,marioo08,ioomar19}).}
\revA{To summarize, at each iteration $j$, one more input of $\mathbf{X_{exp}}$ is added to the covariance function, following the order of inputs in $\mathbf{X_{exp}}$. The $(j-1)$ hyperparameters' values estimated at the $(j-1)\textrm{-th}$ iteration are used as starting points for the optimization algorithm for the $(j-1)$ first hyperparameters. Once all the inputs of $\mathbf{X_{exp}}$ have been included, the last iteration consists in adding $\mathbf{X_{SII}}$ through the isotropic covariance function (see \cite{ioomar19} for the detailed sequential building process).}

Once the Gp hyperparameters are estimated, the Gp is conditioned by the observations of the learning sample to obtain the so-called ``Gp metamodel'', i.e., the resulting conditional random process, which is still a Gp, denoted by $Y_{\textrm{Gp}}(\lbrace \mathbf{X_{exp}} \cup \mathbf{X_{SII}}  \rbrace)$. For each unobserved prediction point, it is therefore fully characterized by its mean and variance (see \cite{raswil06} for an explicit expression of Gp mean and Gp covariance). The conditional mean, denoted by $\widehat{Y}_{\textrm{Gp}}(\lbrace \mathbf{X_{exp}} \cup \mathbf{X_{SII}}  \rbrace)$, is used as a predictor. The conditional variance, denoted by $\operatorname{MSE}[\widehat{Y}_{\textrm{Gp}}(\lbrace \mathbf{X_{exp}} \cup \mathbf{X_{SII}}  \rbrace)]$, is also the mean squared error (MSE) of the predictor. This prediction variance is used to build a confidence interval around the prediction. Covariance between predictions is also available via the conditional covariance.


\subsection{Validation of the Gp metamodel}

The accuracy and prediction capabilities of the Gp metamodel are assessed by ``$K$-fold cross-validation'' \cite{hastib09}, given the limited budget of simulations. Note that the criteria presented in the following can obviously be calculated on a test basis if available. However, for the considered applications, no test sample is available. For $K$-fold cross-validation, rule-of-thumb methods suggest to fix large values of $K$ (typically $5$, $10$ or $20$) since it is usually preferable to exploit a larger number of simulations for training purposes, at the expense of a loss of generalization of the estimated errors.

First, to quantify the accuracy of predictor, one can use the predictivity coefficient $Q^2$: 
\begin{equation}
Q^2=1-\frac{\sum_{i=1}^{n}\left(y^{(i)}-\widehat{Y}^{(i)}_{\textrm{Gp}, -i}\right)^2}{\sum_{i=1}^{n} \left( y^{(i)} - \frac{1}{n}  \sum_{i=1}^{n}  y^{(i)} \right)^2}
\end{equation}
where $y^{(i)}$ and $\widehat{Y}^{(i)}_{\textrm{Gp}, -i}$ are respectively the $i$-th observation of the learning sample and the corresponding prediction of the Gp metamodel built without $y^{(i)}$. $Q^2$ corresponds to the coefficient of determination in prediction, computed by cross-validation on the learning sample. The closer to one the $Q^2$, the better the accuracy of the metamodel. A plot of predicted values against observed values ($\widehat{Y}^{(i)}$ vs. $y^{(i)}$) or a quantile-quantile plot can also be drawn. In the purpose of identifying the input area yielding to critical configurations, we also compute the rate of good prediction of $\{Y>\widehat{q}_{0.9}\}$.

Second, the quality of the Gp prediction intervals should also be evaluated (intervals built from the prediction variance). For this, the histogram of the predicted standardized residuals can be plotted (a standard Gaussian distribution should be observed). 
The Predictive Variance Adequacy (PVA) is also computed: this criterion defined by \cite{bac13}, and more recently used in \cite{demioo21}, assesses whether the prediction errors are of the same order of the prediction variances or not:
\begin{equation}
\operatorname{PVA} \displaystyle = \displaystyle \log\left[ \frac{1}{n} \sum_{i=1}^{n}  \left(\frac{\sum_{i=1}^{n}\left(y^{(i)}-\widehat{Y}^{(i)}_{\textrm{Gp}, -i}\right)^2}{\operatorname{MSE}(\widehat{Y}^{(i)}_{\textrm{Gp}, -i})} \right)\right].
\end{equation}
The smaller the PVA, the more reliable the prediction intervals. For example, a PVA around $0.2$ (respectively $0.7$) corresponds to prediction variances around $20\%$ (respectively two times) too large or too small with respect to the prediction errors.

Finally, we consider a graphical tool introduced by \cite{marioo12}. This tool enables to compare the Gp prediction intervals of level $\alpha$ with respect to the proportions of observations that actually lie within these intervals. These proportions (i.e., the ``observed'' confidence intervals) can be visualized against the $\alpha$-theoretical prediction intervals, for different values of $\alpha$ in $[0,1]$. By definition, the more the points are located around the identity line, the better the adequacy is. This plot will be called ``$\alpha$-$\alpha$ plot'' in the following. For the interested reader, a detailed interpretation of these numerous validation criteria of the Gp metamodel (as well as additional ones, not mentioned in the present work) can be found in \cite{demioo21}.


\subsection{Application on IB-LOCA uses-cases}\label{ssec_Gp_appli}

\begin{figure}[h!]
\begin{center}
	\subfloat[Predicted vs. observed values.]{\includegraphics[width=0.5\textwidth]{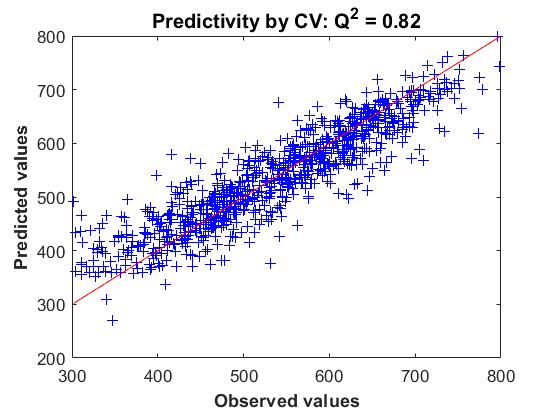}\label{fig:diagnosticPG_IBLOCA1}}
	~
	\subfloat[$\alpha$-$\alpha$ plot.]{\includegraphics[width=0.5\textwidth]{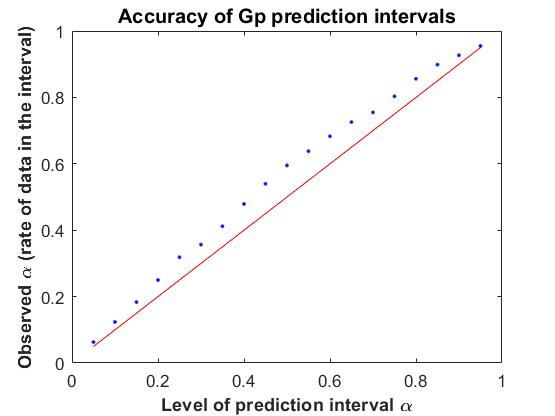}\label{fig:ICPG_IBLOCA1}}		
\end{center}
\caption{IB-LOCA$_1$ use case: diagnostics of Gp performance.}\label{fig:diagnostic_IBLOCA1}
\end{figure}

\begin{figure}[h!]
\begin{center}
	\subfloat[Predicted vs. observed values.]{\includegraphics[width=0.5\textwidth]{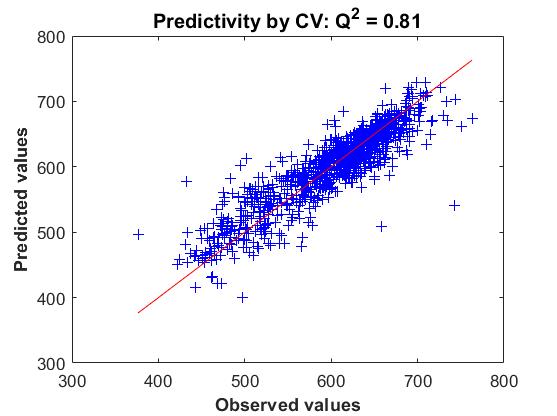}\label{fig:diagnosticPG_IBLOCA2}}
	~
	\subfloat[$\alpha$-$\alpha$ plot.]{\includegraphics[width=0.5\textwidth]{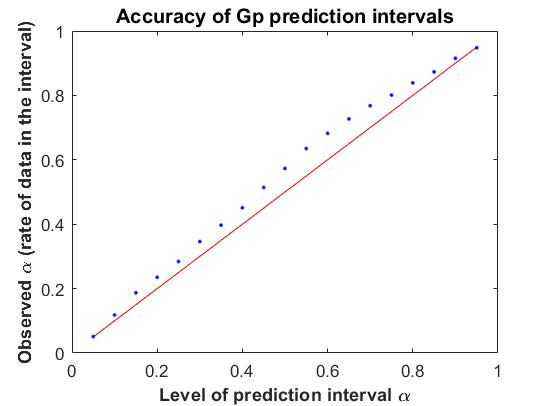}\label{fig:ICPG_IBLOCA2}}		
\end{center}
\caption{IB-LOCA$_2$ use case: diagnostics of Gp performance.}\label{fig:diagnostic_IBLOCA2}
\end{figure}

Some of the aforementioned diagnostic metrics have been computed with $K=10$ folds for the cross-validation procedure and are plotted in Figures~\ref{fig:diagnostic_IBLOCA1} and~\ref{fig:diagnostic_IBLOCA2} for the two IB-LOCA use cases.
In both cases, the predictivity coefficient $Q^2$ is around $0.8$ which can be considered as satisfactory and reasonable regarding both the final objective of the study (i.e., the uncertainty propagation) and the multiple constraints which have to be handled (limited size of the learning sample, high dimension in input, strong nonlinearity of the model). Note that, in the case of poor predictivity performance, the use of Gp-based iterative enrichment strategies can be relevant to add supplementary code simulations. Again, such a strategy is not possible in the context of the present study. Thus, for the two use cases, less than $20\%$ of the output variability remains not explained by the Gp metamodel. More precisely, this includes both the inaccuracy of the Gp and the loss of information attributable to the inputs in $\mathbf{X_\varepsilon}$ which have not been selected in the metamodel building process. \revB{Another possible reason for the part of unexplained output variability can be the behavior of the code itself. For instance, such a locally irregular behavior can be due to physical bifurcations or threshold effects. In our case, a convergence study, for which all the variables in the model were added sequentially (i.e., even the $\mathbf{X_\varepsilon}$), has been performed. Such a study, not presented here for the sake of brevity, showed that no improvement of the predictivity has been detected. It is therefore reasonable to assume that the code behavior also plays a role in this remaining output variability.}
Finally, high rates of good classification for critical configurations are also obtained: $94\%$ and $93\%$ for IB-LOCA$_1$ and IB-LOCA$_2$, respectively.

Note that building the Gp directly in dimension $d = 97$, without variable selection and sequential inclusion processes, leads to a poor estimation of the Gp (typically, the optimization procedure involved in the estimation of the Gp hyperparameters often fails) and yields to a Gp with a poor predictivity. This illustrates the practical interest of the proposed methodology. For comparison purposes, other metamodels such as robust linear regression (e.g., elastic net regularization of \cite{zouhas02}), support vector machines \cite{scholkopf2002learning} or regression trees \cite{brefri84} were tested. All of them yield poor predictivity with $Q^2$ lower than $0.5$. Only a generalized additive model based on smoothing splines, namely the ``Adaptive COmponent Selection and Shrinkage Operator'' (ACOSSO \cite{stobon11}), with a $Q^2$ around $0.7$, showed competitive results (for IB-LOCA$_2$ only) with respect to the proposed Gp-based approach.

Concerning the quality of predictive variance and confidence intervals, low $\operatorname{PVA}$ values ($0.15$ and $0.02$ for IB-LOCA$_1$ and IB-LOCA$_2$, respectively) and $\alpha$-$\alpha$ plots show that the Gp metamodel yields accurate confidence intervals in prediction, although being sometimes too conservative for central values of $\alpha$. The Gp metamodel provides a reliable prediction error: the conditional distribution of Gp is accurate. In conclusion, the Gp metamodel can be confidently used in a Bayesian framework for uncertainty propagation in the next step.

\section{Step 4: Application for the identification of penalizing configurations}\label{sec:step4}

\subsection{Definition and computation of conditional probabilities}

The final goal of the ICSCREAM methodology is to identify the penalizing configurations (i.e., corresponding to a maximal PCT) of the inputs of interest $\mathbf{X_{pen}}$, \revA{under} the uncertainty of the other inputs. To do so, one can define, for any set of possible values of $\mathbf{X_{pen}}$, the probability of exceeding the critical value $\widehat{q}_{0.9}$. This conditional probability can be estimated for a given value $\mathbf{x_{pen}} \in \mathcal{X}_{\textrm{pen}}$ with the Gp-Bayesian approach by:
\begin{eqnarray}\label{eq:proba}
\widehat{P}(\mathbf{x_{pen}}) & = & P[{Y}_{\textrm{Gp}}(\mathbf{X_{exp}},\mathbf{X_{SII}}) > \widehat{q}_{0.9} \ | \mathbf{X_{pen}} = \mathbf{x_{pen}}] \nonumber\\
& = & 1 - \mathbb{E}[ 1_{{Y}_{\textrm{Gp}}(\mathbf{X_{exp}},\mathbf{X_{SII}}) \leq \widehat{q}_{0.9}} | \mathbf{X_{pen}} = \mathbf{x_{pen}} ] \nonumber\\
& = & 1 - \mathbb{E}[ 1_{{Y}_{\textrm{Gp}}(\mathbf{\widetilde{X}_{exp}},\mathbf{X_{pen}}) \leq \widehat{q}_{0.9}} | \mathbf{X_{pen}} = \mathbf{x_{pen}} ] \nonumber\\
& = &  1 - \mathbb{E}\left[  \mathbb{E} \left[ 1_{{Y}_{\textrm{Gp}}(\mathbf{\widetilde{X}_{exp}},\mathbf{X_{pen}}) \leq \widehat{q}_{0.9}} | \mathbf{\widetilde{X}_{exp}} \right] \; | \; \mathbf{X_{pen}} = \mathbf{x_{pen}} \right]  \nonumber\\
& = & 1 - \displaystyle \bigintsss_{\mathcal{\widetilde{X}}_{exp}} \Phi \left( \frac{\widehat{q}_{0.9} - \widehat{Y}_{\textrm{Gp}}(\mathbf{\widetilde{x}_{exp}},\mathbf{x_{pen}})}{\sqrt{\operatorname{MSE}[ \widehat{Y}_{\textrm{Gp}}(\mathbf{\widetilde{x}_{exp}},\mathbf{x_{pen}})]}}\right) d\mathbb{P}_{\mathbf{\widetilde{X}_{exp}}} (\mathbf{\widetilde{x}_{exp}})
\end{eqnarray}
where $\mathbf{\widetilde{X}_{exp}} =\lbrace \mathbf{X_{exp}}\cup \mathbf{X_{SII}}\rbrace\setminus \mathbf{X_{pen}}$ denotes the Gp inputs deprived of $\mathbf{X_{pen}}$, $\mathcal{\widetilde{X}}_{\textrm{exp}}$ their domain of variation, $d\mathbb{P}_{\mathbf{\widetilde{X}_{exp}}}$ their probability density function, and $\Phi(\cdot)$ the cumulative distribution function of the standard Gaussian distribution. Note that the independence between $\mathbf{\widetilde{X}_{exp}}$ and $\mathbf{X_{pen}}$ is required and used to obtain the fourth line from the third one in  Eq.~(\ref{eq:proba}).
\revA{In the so-called ``Gp-Bayesian'' approach, the predictive distribution of the Gp is entirely propagated (including the Gp error prediction); it is strictly equivalent to a calculation by conditional simulations. This approach therefore differs from a ``plug-in'' approach where only the Gp predictor (i.e., the Gp mean) is considered.
A comparison between the Gp-Bayesian and plug-in approaches is given for example in \cite{ioomar19} (for a quantile estimation problem).}
    
In practice, this conditional probability can be computed for the different values of each input in $\mathbf{X_{pen}}$, as well as for any subgroup of $l$ inputs ($l=2,\ldots,\mbox{dim}(\mathbf{X_{pen}})$). Thus, it results in a one (or $l$)-dimensional function. 

\subsection{Results for IB-LOCA use cases}

\begin{figure}[!ht]
\centering
\includegraphics[width=\textwidth]{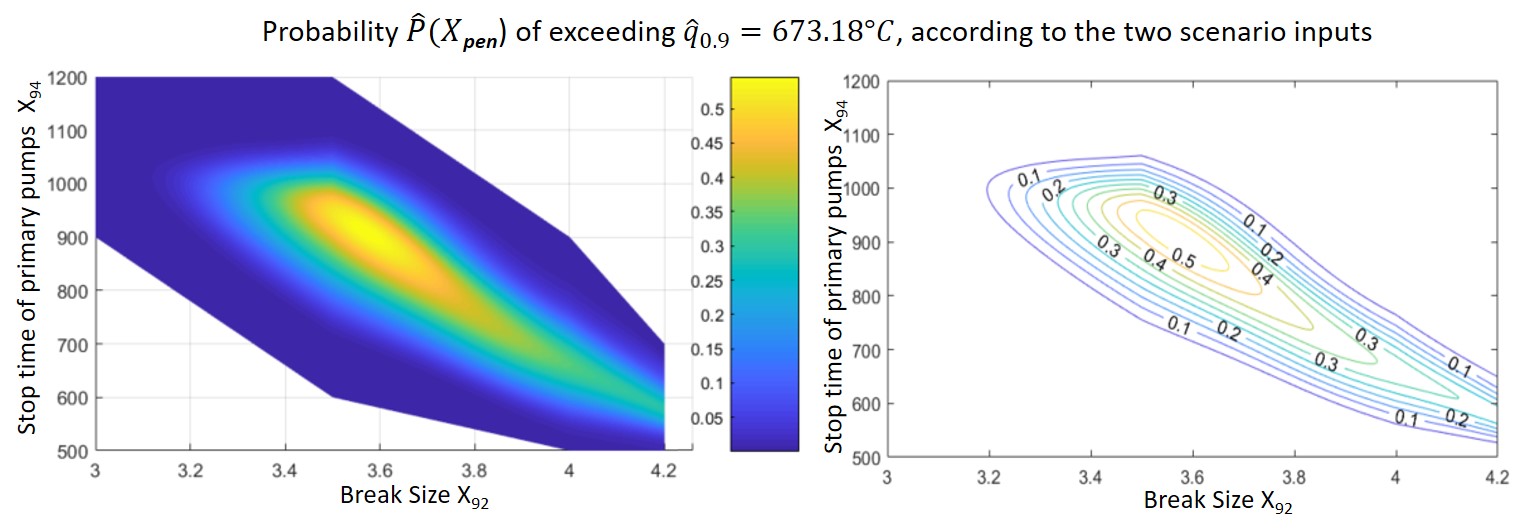}
\caption{\label{fig:IBLOCA1_cond_proba}IB-LOCA$_1$ use case: conditional probability $\widehat{P}(X_{92}, X_{94})$ of exceeding $\widehat{q}_{0.9}$, estimated from the Gp, with both surface and contour plot representations.}
\end{figure}

\begin{figure}[!ht]
\centering
\includegraphics[width=\textwidth]{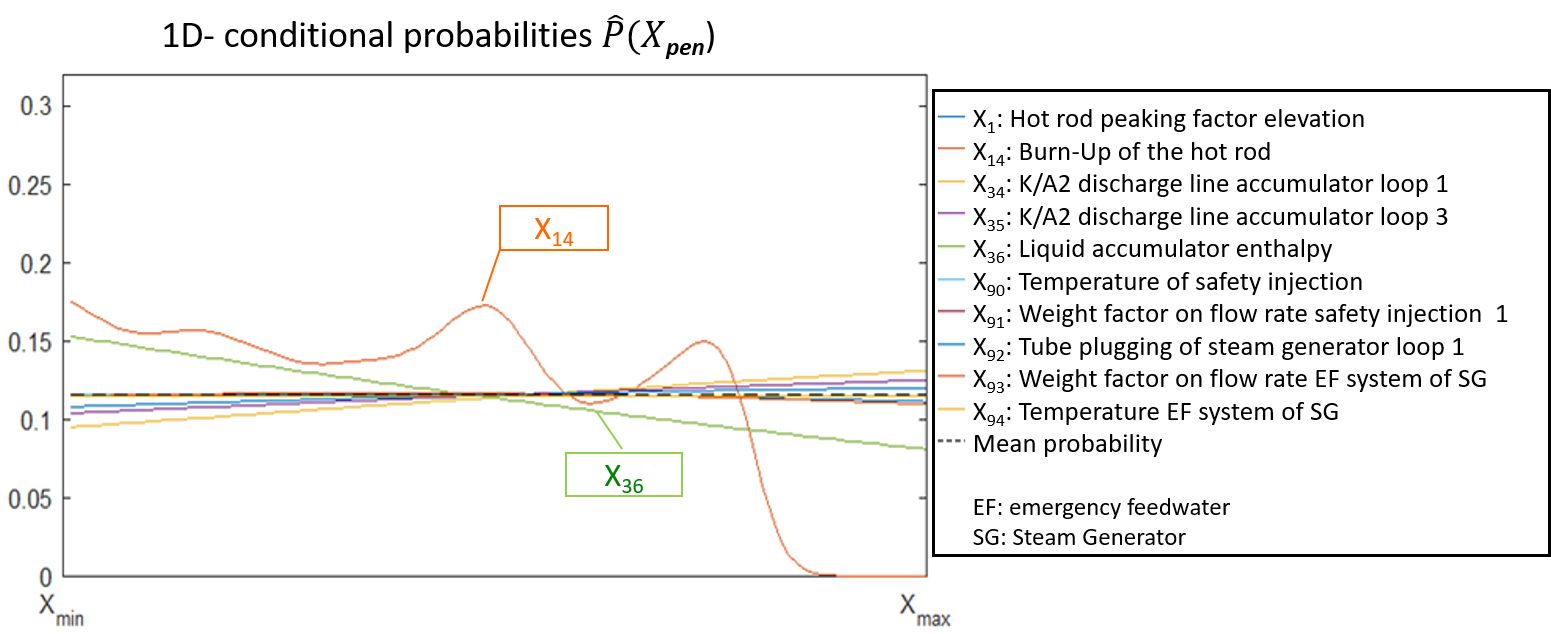}
\caption{IB-LOCA$_2$ use case: conditional probabilities of exceeding $\widehat{q}_{0.9}$ for the ten inputs to be penalized $\mathbf{X_{pen}}$, estimated from the Gp on their respective ranges of variation $[X_{min} ; X_{max}]$. The mean probability $P[Y(\mathbf{X_{exp}},\mathbf{X_{SII}}) > \widehat{q}_{0.9} ]$ is also estimated to $0.118$ and plotted in black dotted line.}\label{fig:IBLOCA2_cond_proba}
\end{figure}

The conditional probabilities given by Eq.~(~\ref{eq:proba}) are, in practice, computed by intensive crude Monte Carlo Gp-based simulations with respect to the couple $\left( X_{92}, X_{94} \right)$ in IB-LOCA$_1$ and for each input of $\mathbf{X_{pen}}$ in IB-LOCA$_2$. \revB{Here, we use $2\times10^4$ Monte Carlo simulations to estimate each probability. Note that, for a target probability to be estimated close to $0.1$, such a budget corresponds to a coefficient of variation of $2\%$ approximately. Consequently, it is more than enough (especially considering the accuracy of the metamodel itself). The satisfactory accuracy of the estimated probabilities can also be verified by a convergence graph as the number of Monte Carlo Gp simulations increases.} The estimated probabilities are given in Figures~\ref{fig:IBLOCA1_cond_proba} and~\ref{fig:IBLOCA2_cond_proba}, respectively.

For the IB-LOCA$_1$ use case, the analysis of conditional probabilities reveals the strong interaction of the two scenario inputs ($X_{92}$ and $X_{94}$) in the occurrence of critical configurations. The worst cases (i.e., where the probability reaches maximal values) correspond to medium values of both inputs. Conditionally to a given break size ($X_{92}$), the probability function according to the stopping time of primary pumps ($X_{94}$) has a bell shape that reaches its maximum value for a stopping time which decreases linearly as the break size increases. Note that the worst configuration is obtained for a break size equal to $3.57$ inches and a stopping time of the primary pumps of about $907.8$ seconds, which leads to a probability of exceeding the quantile $\widehat{q}_{0.9}=673.18^{\circ}$C estimated at $0.55$.
These results can be enlightened by an analysis of the physical phenomenon. In a nutshell, the correlation between these two scenario inputs drives the degradation of the water inventory. The smaller the break size, the longer the pump will have to run for the same inventory degradation. As for the distinct left and right limits on the domain, they can also be explained. On the one hand, if $X_{92} <3.3$ inches, meaning that the break size is rather small, the water inventory does not degrade too much (whatever the primary pump does). This leads to a slow LOCA which can be contained by the protection systems which have enough time to intervene (hence, the net border). On the other hand, when the break size increases too much, the break tends to be prevailing and reduces the impact of the stop time of the primary pumps (hence, the fading area).

For the IB-LOCA$_2$ use case, amongst the ten scenario inputs, only two are significantly influential on the probability of exceeding $671.85^{\circ}$C. 
The enthalpy of the liquid accumulator ($X_{36}$) has a moderate linear effect: the probability is increased by $0.03$ for the lower bound of $X_{36}$ (penalizing value) and decreased by the same amount for the upper bound. This effect is consistent with the physical knowledge related to the impact of this input on the condensation effect during the safety injection.
More interestingly, the strong, complex and non-monotonic effect of the burn-up is clearly illustrated: the probability of PCT is maximal for the lower bound ($\widehat{P}(x_{14,\text{min}})=0.175$) then decreases before increasing to reach a local maximum value around $x_{14} \simeq 29000$ MWd/t, almost as high as the global maximum value (the variation range of this input is $[515;59000]$). It then decreases prior to increasing until another smaller peak, and then falls to a zero probability of exceeding $\widehat{q}_{0.9}$, for the highest values of $X_{14}$. The burn-up, which is an integral measure of how much energy is extracted from a primary nuclear fuel source, describes the fuel depletion as it supplies energy. When the fuel is too depleted (values of burn-up $ \geq 50000$ MWd/t), the residual power significantly decreases as the burn-up increases, yielding to low PCT values. On the contrary, for lower depletion rate ($X_{14} < 50000$ MWd/t), the relationship between the burn-up and PCT is much more complex since this input controls a large number of other physical quantities related to the life of the hot rod.
\revB{Therefore, the implication of the burn-up in several physical sub-models (as the pressure inside the cladding, the initial temperature of the fuel and the neutronic data), constituent of the global thermal-hydraulic model, explains the form of $\widehat{P}(\mathbf{X_{14}})$: the first small bumps are due to fluctuations in fuel temperature over the life of the rod, while the large drop at the end is due to the decay of the neutronic data at the end of the rod lifetime.}

From a statistical methodology point of view, it is really interesting to clearly highlight some behavior, which are known by physicists, but impossible to detect on the initial dataset by simple visualization (cf. scatter plots of Figure~\ref{fig:MDTE_scatter_scenario}).
The results obtained with the proposed ICSCREAM methodology are therefore consistent with physical knowledge and even go beyond this. Indeed, they enable to quantify the impact on the probability but also to precisely identify and justify the penalizing areas in input.

\section{Conclusion and prospects}\label{sec:ccl}

In the framework of risk assessment in nuclear accident analysis, it is essential to quantitatively assess the uncertainties tainting the results of best-estimate computer codes. Beyond the usual uncertainty propagation step, this paper has focused on identifying the penalizing configurations of some specific inputs (the scenario ones), \revA{under} the uncertainty of the other inputs. This methodology, called ICSCREAM, was motivated by a study at the reactor-scale of an IB-LOCA scenario in a pressurized water reactor, with the thermal-hydraulic CATHARE2 code. In the considered use cases, around a hundred scalar inputs are uncertain and the penalizing values of two or ten of them must be identified. The output variable of interest is the peak cladding temperature (PCT) during the accident transient. The critical configurations are defined by PCT exceeding its quantile of probability $0.9$. 
The fidelity and complexity of the numerical modeling, the limited budget of simulation and the very large number of uncertain inputs are real challenges that led to the development of a sophisticated and robust methodology based on advanced statistical tools.

Applied from a single Monte Carlo sample of CATHARE2 simulations, the ICSCREAM methodology judiciously combines a step of sensitivity analysis to identify and rank the main influential inputs and to reduce the dimension, with a \revB{specific building procedure (based on a sequential inclusion of variables)} of a Gaussian process (Gp) metamodel. The sensitivity analysis step relies on new statistical independence tests which aggregate information of global and target Hilbert-Schmidt independence criterion (HSIC) measures. From the sensitivity analysis results, the inputs are taken into account in a more or less fine way in the Gp metamodel, according to their supposed influence. The non-selected inputs are integrated in the variance of the Gp prediction. The robustness of the metamodel is therefore enhanced and its building is made possible in such a high-dimensional problem. Once built, the accuracy and prediction capabilities of the Gp metamodel are assessed: several graphical and quantitative tools are proposed for this purpose. Finally, the Gp metamodel can be used intensively for uncertainty propagation, estimation of conditional probabilities and for the inversion problem. The conditional probabilities of PCT exceeding the critical value are here estimated using the Gp within a Bayesian framework to take into account the uncertainty of the Gp prediction.

The efficiency of ICSCREAM methodology has been demonstrated on two IB-LOCA use cases of high complexity and from a learning sample of around a thousand simulations. For both cases, $80\%$ of the PCT variance is explained by the Gp, with a high ability for identifying PCT critical areas. Although the metamodel could still be improved, it provides a reliable prediction error and accurate prediction intervals. This justifies its use in a Bayesian framework for the estimation of conditional probabilities of exceeding the $90\%$-quantile, according to each input (or group of inputs) that have to be penalized. This estimation, not directly feasible with the numerical thermalhydraulic model because of its computational cost, becomes now tractable via intensive simulations of the Gp metamodel. 
The analysis of the conditional probabilities has revealed the strong and non-monotonic individual influence of some scenario inputs for one use case, and the high interaction of the two scenario inputs for the other one.

ICSCREAM has been developed in a particular industrial context where the user cannot add new simulations to the current design of experiments.
In the case where additional simulations could be available, a research track would be to develop dedicated adaptive sampling strategies.
Following this idea, efficient Gp-based sequential enrichment schemes could be adapted to add code simulations in an efficient manner \revA{(see, e.g., the works of \cite{becgin12,chebec14,moustapha:2016,wang:2020,sabater:2021} which deal with various purposes)}. However, their direct use in such large dimensional problems still remains a challenge. Another important limitation of the ICSCREAM methodology is its use on large-size datasets, typically several thousands of simulations. The estimation and building of the Gp metamodel then become difficult to achieve and compute. To make the Gp applicable in this context, methods based on nested kriging \cite{ruldur18} or variational inference \cite{henfus13} could be studied. Finally, handling the ``chaotic'' code behavior (e.g., physical bifurcations or threshold effects) could be achieved by non-stationary Gp metamodels such as the treed Gp which allows a smart domain partitioning \cite{gralee08}.

\revA{\section*{Appendix -- Illustration on an analytical test case}\label{sec:analytic}
Initially, the ICSCREAM methodology was directly developed to be applied to very high-dimensional real datasets. To further illustrate the robustness and efficiency of the method, we propose here an additional analytical test case, in lower dimension, but applying the same rules of thumb for the sample size. This case is designed to be ``representative'' of some features of the underlying physics at stake in an IB-LOCA model, in the sense that it exhibits interaction and non-monotonic conditional probabilities. This case is proposed to bring additional information to the previous applications by illustrating how a predictivity around $80\%$ allows nevertheless to correctly estimate the conditional probabilities. Moreover, it can be seen as a tool for future research prospects, for example on adaptive sampling in large dimension.

The considered analytical model $\mathcal{M}_{f}$, inspired from the Friedman function \cite{fri91} and defined in dimension $d_f=20$, is given by:
\begin{equation}\label{eq:M_f}
 \mathcal{M}_{f}(\mathbf{X}) =a_1 \sin{\left(6\pi \; X_1^{5/2}(X_2-0.5)\right)}  + a_2 (X_3-0.5)^2 + a_3 X_4 + a_4 X_5 + r_{X6, \ldots, X_{15}}
\end{equation}
where $r_{X_6, \ldots, X_{15}} = \frac{a_5}{\sqrt{(\sum_{i=6 \ldots 15} i^2)}}\sum_{i=6 \ldots 15} \sqrt{12} i (X_i - 0.5)$ and $\mathbf{X}=(X_1,\dots,X_{20})$ are independent and uniform random variables on $[0,1]$.

The model depends only on the fifteen first inputs. The first term represents a strong and non monotonic interaction between the two first inputs. The second term is a quadratic function of $X_3$ while the other ones are linear. The inputs that have to be penalized are ($X_1,X_2,X_3,X_4,X_{5})$. The parameters for tuning the influence of the different inputs are chosen as follows: $\mathbf{a} = (5, \; 20, \; 8, \; 5, \; 1.5)^\top$. 
Under this parametrization, $X_2$ explains alone around $10\%$ of the output variance, $X_1$ has no individual effect but its interaction effect with $X_2$ is strong (around $30\%$ of the output variance). 
$X_3$, $X_4$ and $X_5$ only have individual effects (no interaction) and explain, respectively, around $11\%$, $28\%$ and $10\%$ of the output variance.
The effects of the ten remaining inputs ($X_6$ to $X_{15}$) represent around $11.5\%$ of the output variance. 
Finally, we compute reference values (by intensive Monte Carlo simulations) of the $90\%$-quantile of the model output (approximately equal to $q_{0.9} = 14$) and the one-dimensional mean effects as well as the one-dimensional conditional probabilities of exceeding $q_{0.9}$, according to the inputs to be penalized (see Fig.~\ref{fig:Mf_theo_E_cond_proba}).

\begin{figure}[!ht]
\centering
\includegraphics[width=\textwidth, height = 6cm]{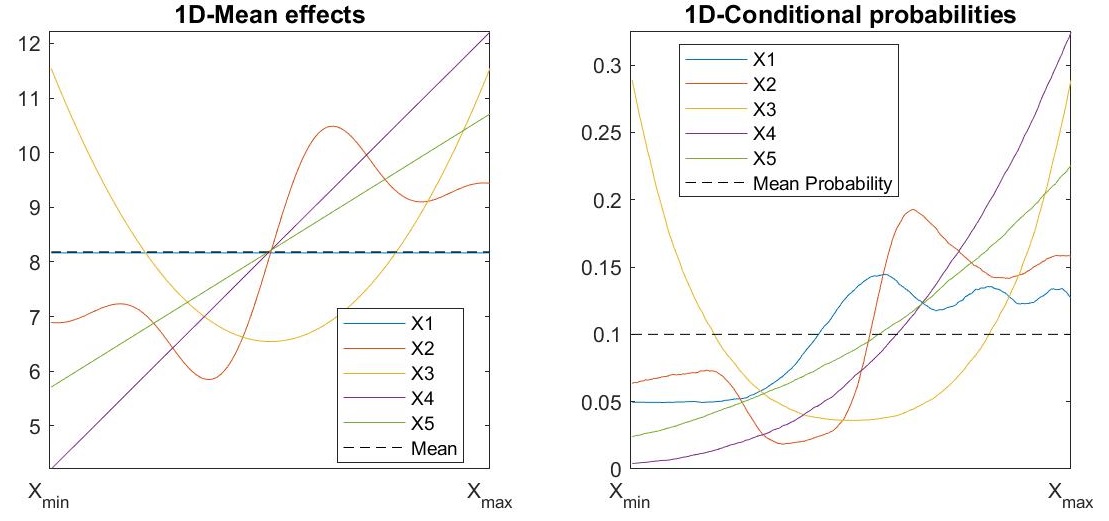}
\caption{$ \mathcal{M}_{f}$ analytical case: reference one-dimensional mean effects and conditional probabilities of exceeding $q_{0.9}$ for the five inputs to be penalized $\mathbf{X_{pen}}$.}\label{fig:Mf_theo_E_cond_proba}
\end{figure}

To apply the ICSCREAM methodology, we consider a Monte Carlo sample of $n=200$ evaluations of the $\mathcal{M}_{f}$ model and the $90\%$-quantile is estimated from this learning sample ($\widehat{q}_{0.9} = 13.8$ is obtained here). Step~$2$ is performed with a level $\alpha_{\textrm{test}} = 0.05$ which yields to the selection of inputs $X_1$ to $X_5$ by HSIC measures and $X_1$ to $X_6$ by target HSIC ones. From an aggregation procedure based on Bonferroni’s correction, $X_1$ to $X_5$ are selected in $\mathbf{X_{PII}}$ and $X_6$ in $\mathbf{X_{SII}}$. From this selection, a Gp metamodel is built with the same parametric choices as in Section \ref{ssec_Gp_appli}. Predictivity coefficients of $Q_2 = 0.78$ and $Q_2 = 0.77$ are obtained on a test basis and by cross-validation on the learning sample, respectively. Regarding the predictive variance, $PVA$ is around $0.15$. Diagnostic metrics of the Gp performance validation are illustrated in Figure \ref{fig:diagnostic_Mf}. We observe that, despite the amount of unexplained variance (i.e., slightly more than $20\%$), the Gp prediction intervals are satisfactory. We are therefore in a configuration quite similar to those obtained on the two IB-LOCA use cases.

From the Gp metamodel, the conditional probabilities are estimated and provided in Figure \ref{fig:Mf_PG_E_cond_proba}. We observe a good approximation of the one-dimensional conditional probabilities (by comparing to the reference curves given in Fig.~\ref{fig:Mf_theo_E_cond_proba} right): first, the different behaviors are well captured; second, the penalizing areas of each input are clearly identified. 
Finally, a convergence study has also been performed: it shows the predictivity and Gp performance improvements as $n$ increases ($Q^2$ around $0.85$ and $0.94$, for $n = 400$ and $600$ respectively). 

In the future, such a test case can be extended by adding, for instance, some \revB{locally irregular} behavior or an interaction between an input to be penalized $X_A$ and another input $X_B$ (not to be penalized) whose global influence would be rather small on the output but significant on the conditional probability to $X_A$. In this last case, the risk would be that $X_B$ would not be selected by HISC-based tests (e.g., if the sample size is too small), which would lead to a poor estimation of the conditional probability of $X_A$. The challenge will be to be able to detect this type of error, perhaps by relying on convergence and stability plots of HSIC, or by building more powerful HSIC-based tests. Finally, the modeling and propagation of the ``observed'' error of the metamodel could be investigated and compared to the predictive error of the metamodel.

\begin{figure}[h!]
\begin{center}
	\subfloat[Predicted vs. observed values.]{\includegraphics[width=0.5\textwidth]{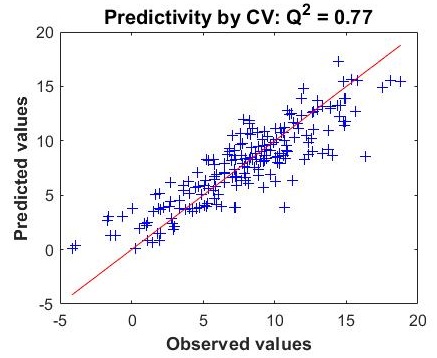}\label{fig:diagnosticPG_Mfa}}
	~
	\subfloat[$\alpha$-$\alpha$ plot.]{\includegraphics[width=0.5\textwidth]{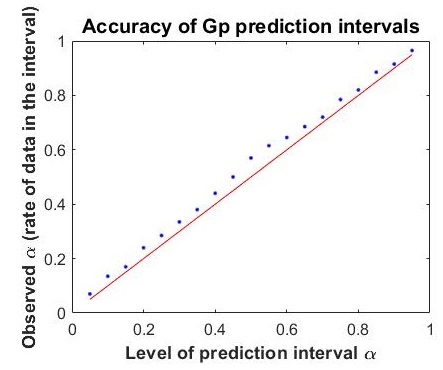}\label{fig:diagnostic_Mfb}}		
\end{center}
\caption{$ \mathcal{M}_{f}$ analytical case:  diagnostics of Gp performance.}\label{fig:diagnostic_Mf}
\end{figure}

\begin{figure}[!ht]
\centering
\includegraphics[width=0.6\textwidth]{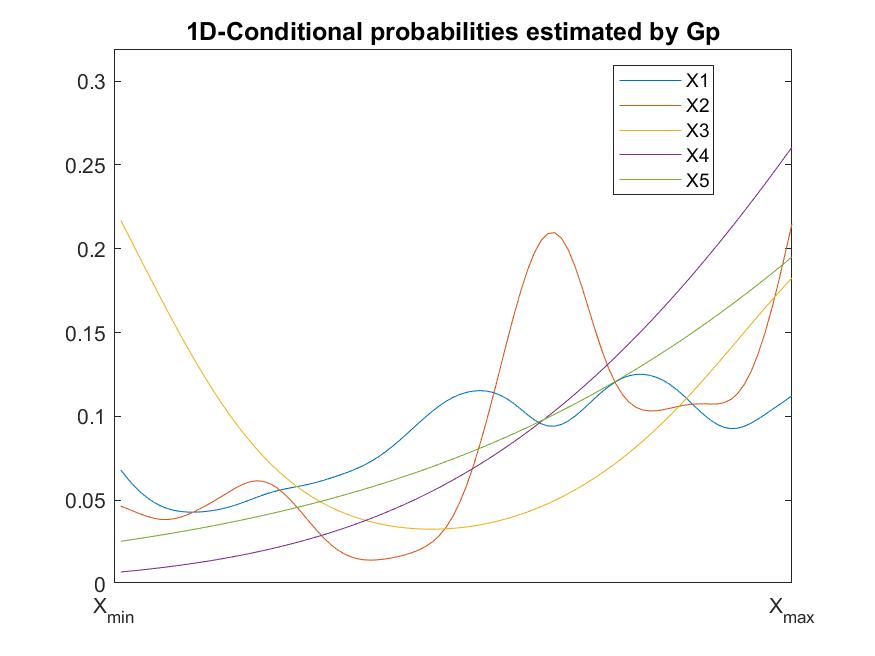}
\caption{$ \mathcal{M}_{f}$ analytical case: conditional probabilities of exceeding $\widehat{q}_{0.9}$ for the inputs to be penalized, estimated from the Gp.}\label{fig:Mf_PG_E_cond_proba}
\end{figure}
}

\section*{Acknowledgments}
The writing of this paper was partly funded by the ANR project SAMOURAI (ANR-20-CE46-0013).
We warmly thank our two reviewers whose remarks have greatly helped to improve the paper.
We are also grateful to Vincent Larget, Mathieu Segond, Laurent Lefebvre and Jean-Christophe Lecoy for numerous discussions about this work. CATHARE2 code is developed under the collaborative NEPTUNE project, supported by CEA, EDF, Framatome and IRSN. 


\end{document}